\RequirePackage[2020-02-02]{latexrelease}

\documentclass[fleqn]{jaes}

\jyear{2023}
\jmonth{July}
\makeatletter
\renewcommand\paragraph{\@startsection{paragraph}{4}{\z@}%
            {-2.5ex\@plus -1ex \@minus -.25ex}%
            {1.25ex \@plus .25ex}%
            {\normalfont\normalsize\bfseries}}
\makeatother
\usepackage{amsmath}
\usepackage{bm}

\usepackage{hyperref}
\usepackage{cleveref}
\usepackage[hypcap=true]{caption}

\usepackage{booktabs}
\usepackage{multirow}

\begin{document} 
% Page heads
%\markboth{-}{JAES TEMPLATE}
%\markboth{Vanka and Safi}{JAES TEMPLATE}

% Title portion
\title{The Role of Communication and Reference Songs in the Mixing Process: Insights from Professional Mix Engineers\thanks{To whom correspondence should be addressed, e-mail: s.s.vanka@qmul.ac.uk.}}

%Author Info.
\authorgroup{
\author{Soumya Sai Vanka\textsuperscript{1}},
\author{Maryam Safi\textsuperscript{2}},
\author{Jean-Baptiste Rolland\textsuperscript{2}},
AND 
\author{Gy\"orgy Fazekas\textsuperscript{1}}
\email{\quad(s.s.vanka@qmul.ac.uk)\quad\quad\quad\quad(m.safi@steinberg.de)\quad\quad\quad\quad(jb.rolland@steinberg.de)\quad\quad\quad(george.fazekas@qmul.ac.uk)}
\affil{\textsuperscript{1}Centre for Digital Music, Queen Mary University of London, London, UK\\
\textsuperscript{2}Steinberg Media Technologies GmbH, Hamburg, Germany}
}

%----------------------------
\abstract{%
Effective music mixing requires technical and creative finesse, but clear communication with the client is crucial. The mixing engineer must grasp the client's expectations, and preferences, and collaborate to achieve the desired sound. The tacit agreement for the desired sound of the mix is often established using guides like reference songs and demo mixes exchanged between the artist and the engineer and sometimes verbalised using semantic terms. This paper presents the findings of a two-phased exploratory study aimed at understanding how professional mixing engineers interact with clients and use their feedback to guide the mixing process. For phase one, semi-structured interviews were conducted with five mixing engineers with the aim of gathering insights about their communication strategies, creative processes, and decision-making criteria. Based on the inferences from these interviews, an online questionnaire was designed and administered to a larger group of 22 mixing engineers during the second phase. The results of this study shed light on the importance of collaboration, empathy, and intention in the mixing process, and can inform the development of smart multi-track mixing systems that better support these practices. By highlighting the significance of these findings, this paper contributes to the growing body of research on the collaborative nature of music production and provides actionable recommendations for the design and implementation of innovative mixing tools.
}
\maketitle 
%Head 1
\setcounter{section}{0}
\section{Introduction}

\subsection{Music Production and Mixing}

Music production encompasses the entire life cycle of the creation of a song, from idea conception to distribution. The process involves writing the score, recording vocals and instruments, mixing the music, mastering the audio, and finally distributing it~\cite{burgess2013art} that influences the creation of different edits or versions for different formats and media. Audio recordings are made in a studio using digital audio workstations (DAW) and various other equipment. A mixing engineer then combines these recordings using knowledge of signal processing, sound, and music to produce a cohesive and well-balanced mix which is further mastered and enhanced for distribution.

% \subsection{Multitrack Music Mixing}
Music mixing is the process of combining various audio recordings to create a polished, balanced, and artistically aligned finished product that elicits feelings and tells a story ~\cite{miller2016mixing}. The mixing engineer adjusts technical and artistic elements, emphasising crucial components, balancing instrumentation, and employing techniques like gain staging, equalisation, panning, compression, and modulation effects to shape the sound ~\cite{izhaki2017mixing}.

\subsection{Technological Advancements and the Recording Industry}
Technological advancements have transformed the landscape of music production. Multi-track recording, analogue synthesizers, DAWs, and software plugins have given artists and producers greater control, resulting in more intricate and detailed productions~\cite{hracs2012creative}. However, this has led to specialised roles such as recording engineer, producer, mixing engineer and mastering engineer, requiring extensive skills and knowledge~\cite{burgess2014history}. The increased number of stakeholders in the production process demands teamwork and effective communication to achieve the desired goals~\cite{wilsmore2022coproduction}.

% \subsection{The Democratisation of Music Production}
In the past, professional music production required access to expensive studios and equipment. However, affordable recording equipment and software have democratised music production, allowing artists to create high-quality music at home. This accessibility has led to a wider variety of styles and genres, catering to different user groups~\cite{sandler2019semantic,mcgrath2016making, bromham2016can}. Recent literature identifies three user types: amateurs, who are typically newcomers to the field without formal training~\cite{hoare2014coming}, pro-ams, who are often skilled individuals without full professional support~\cite{prior2018new, stebbins1977amateur, leadbeater2004pro}, and professionals, who are typically highly skilled paid engineers with industry expertise~\cite{bromham2016can}.
% \subsection{Higher Expectations for Production Quality}
With high-fidelity recordings and polished productions becoming the norm, artists face pressure to create sonically impressive music~\cite{katz2010capturing}. Streaming platforms and digital distribution have intensified competition, prompting musicians to invest in complex production techniques for a unique and captivating sound. Production quality plays a significant role in gaining traction, yet there remains a gap between the accessibility of production tools and the expertise required to use them effectively.

\subsection{Intelligent Music Production and Automatic Mixing}

Intelligent music production systems assist creators in navigating the complex tasks of audio production. These systems provide insights, suggestions and even full automation of control parameters in multitrack projects ~\cite{IMPbook19}. They benefit both amateurs and professionals, offering assistance and streamlining the process ~\cite{moffat19appr, de2017ten}.

% \subsubsection{Automatic Mixing}
Automatic mixing employs algorithms and intelligent systems to automate the audio mixing process~\cite{perez2011automatic}. It adjusts key audio parameters such as volume, panning, equalisation, compression, and spatial effects. Through the analysis of audio signals and the application of predefined rules or machine learning models, automatic mixing algorithms aim to create balanced and cohesive mixes. Various approaches have been explored~\cite{de2017ten, steinmetz2022automix}, including knowledge-engineered~\cite{de2013knowledge}, machine learning-based~\cite{moffat2019machine}, and deep learning-based methods~\cite{steinmetz2020learning, martinez2022automatic, martinez2021deep, koo2022music}. Most of these systems take raw tracks or stems as input and generate a mix as output. 

% \subsubsection{Limitations}
While these approaches provide solutions within limits, they often lack user control to specify context and expectations. As music mixing allows for multiple possible mixes from the same stems or tracks, context plays a crucial role in determining the best fit for the client's needs~\cite{lefford2021context}. Professionals and pro-ams have shown less interest in completely automatic services due to the generic nature of these systems~\cite{vanka2023adoption, tsiros2020towards}. They prefer technology that allows for more control and contextual understanding to achieve desired results~\cite{lefford2021context}. 

\subsection{Collaborative Practice-Informed Design for Smart Mixing Tools}

Creating music involves multiple stakeholders who contribute their expertise to shape the final product. Clear communication and understanding between artists, recording engineers, mixing engineers, producers, and mastering engineers are crucial to realising the artist's vision. To design mixing systems that facilitate effective collaboration, our work explores the communication dynamics between artists and mixing engineers. These findings can inform the development of future smart mixing tools.

During the mixing process, the mixing engineer employs various methods to understand the artist's expectations. The artist uses semantic terms, verbal instructions, reference songs, and demo mixes to convey their needs. These media allow for developing a tacit agreement between the client and the mixing engineer about the context. The decisions made by the mixing engineer greatly impact the clarity, balance, spatialisation, emotion, and dynamics of the song. Collaborative work between the mixing engineer and artist is vital to achieving the desired results while incorporating their respective identities into the music. Our objective is to understand the communication process enabling this vital synergy, and explore how this knowledge might influence the design of smart mixing tools. 

This study utilises semi-structured online interviews with five experienced mix engineers, followed by a questionnaire-based study to validate the findings. Through qualitative data analysis, we explore how clients express their expectations and how mixing engineers interpret and derive insights from these communication channels. The results of this study aim to inspire an interaction-centred approach to designing mixing tools.

\begin{figure*}[th]
    \centering
    \includegraphics[scale = 0.065]{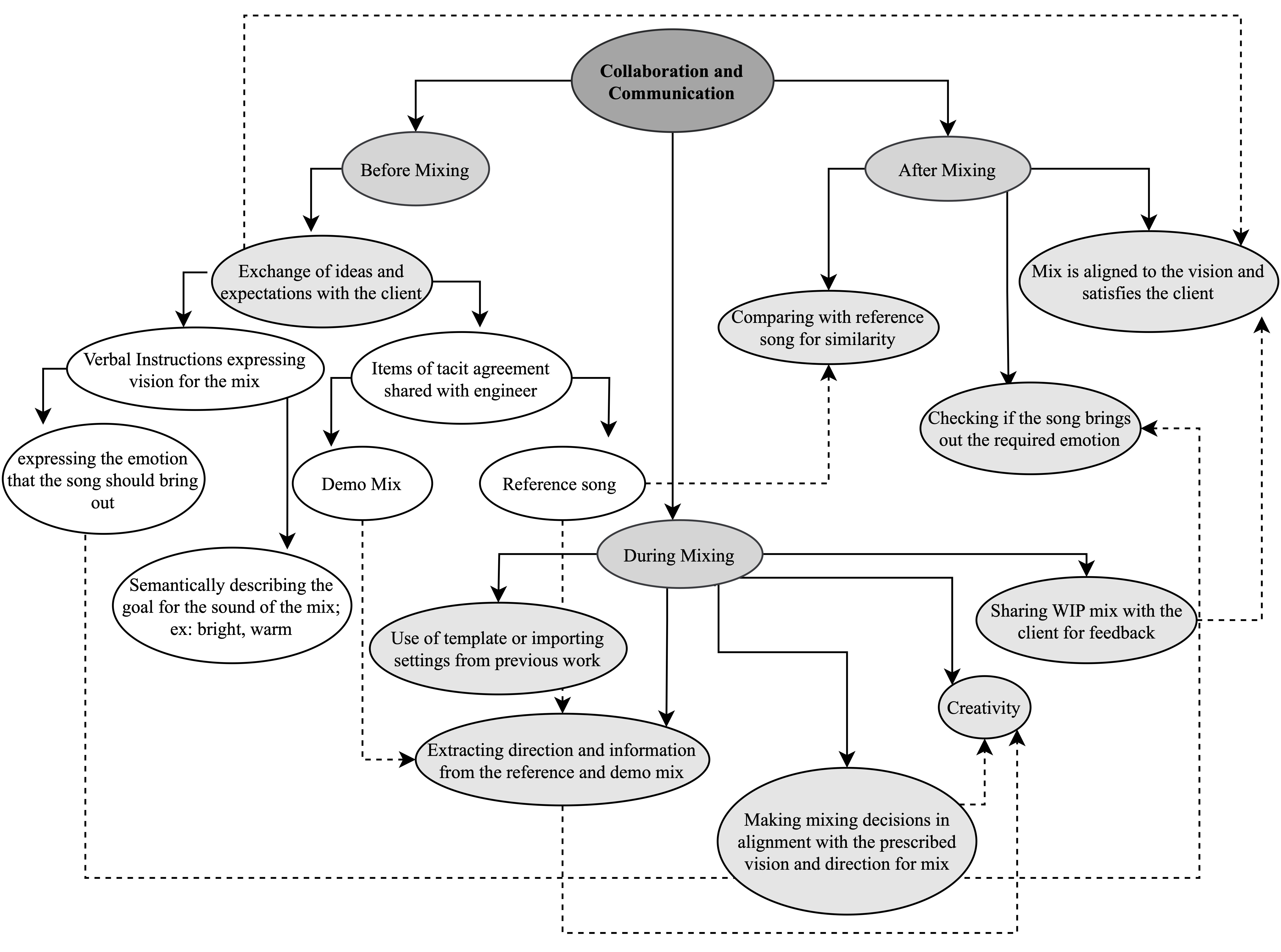}
    \caption{Themes and sub-themes extracted from the thematic analysis for the interviews}
    \label{fig:themes}
\end{figure*}

\section{Study Design and Methodology}
The objective of this study is to investigate the exchange of ideas and expectations between clients and mixing engineers during the song-mixing process. We aimed to understand the informative cues and suggestions provided by clients to convey their objectives for the mix, how mixing engineers interpret and incorporate these cues into the mixing decisions, as well as how mixing engineers assess the completion of a mix. To achieve these goals, we implemented a multi-phase approach consisting of semi-structured interviews and a structured web-based questionnaire.

\subsection{Phase 1: Semi-Structured Interviews}
Semi-structured interviews are qualitative research methods that involve asking a pre-determined set of open questions, allowing for further exploration of responses~\cite{kallio2016systematic}. In this phase of the study, we conducted one-to-one semi-structured interviews with professional mixing engineers to gain insights into their experiences and perspectives.

\subsubsection{Interview Structure}
The interviews focused on exploring various aspects of communication, expectation exchange, and collaboration between mixing engineers and artists during the music-mixing process. Key themes explored during the interviews included the interaction with clients, the role of demo mixes and reference songs, the criteria for defining the completion of a mix, and the role of creativity in the mixing process. The interviews were conducted through recorded online video calls, and the recordings were later transcribed for analysis purposes.

\subsubsection{Participant Profile}
To ensure the validity of the investigation and mitigate selectivity issues, five professional mixing engineers were invited for interviews~\cite{bogner}. By involving multiple experts in the field, we wanted to gain diverse perspectives and minimize the risk of bias or individual preferences from skewing the results. These engineers were chosen based on their expertise and extensive experience in the field of music mixing. At the time of the interviews, they all had more than five years of experience, possessed advanced mixing skills, and actively worked on music projects for themselves and other artists. Their qualifications included formal training, payment for their work, and having their work published using traditional discographies and music streaming platforms. Two of the interviewed engineers are female. Three out of five engineers worked in Europe, one in Asia, and one in America/Europe. Three of them specialise in pop, rock, and metal with the other two mixing for jazz, ensemble, folk, classical and world music.

\subsubsection{Data Analysis}
The interview recordings were transcribed, and thematic analysis was conducted using an inductive, iterative, and grounded approach~\cite{braun2012thematic} ~\cite{braun2006using} in the MAXQDA software \footnote{https://www.maxqda.com/ }. A series of codes were generated to identify different themes across the interviews, which were refined over several iterations. These themes were then used to inform the design of the questionnaire. This involved identifying key themes, patterns, and categories of information related to the exchange of ideas and expectations between the client and the mixing engineer. The analysis process focused on extracting meaningful insights and understanding the underlying factors influencing the mixing process. \autoref{fig:themes} provides a concise summary of the extracted themes and sub-themes, highlighting their interconnectedness. The central theme revolves around the critical role of communication and collaboration throughout the mixing process. Furthermore, we have identified three distinct stages where this communication and collaboration occur: prior to mixing, during the mixing process, and upon completion of the mix. \autoref{fig:themes} effectively illustrates how various sub-themes within different categories intertwine, ultimately forming a comprehensive understanding of the subject.

\subsection{Phase 2: Questionnaire-based Study}
After the analysis of interviews, a structured web-based questionnaire was designed to both validate and expand upon the insights gathered. The qualitative analysis of the interview transcripts allowed us to structure the inferences into questions that could be asked to a wider group of mixing engineers for greater validity.

For example, one key inference from the interviews was that mixing engineers often utilize reference songs from genres different from the ones they are working on. To explore this further, we included specific questions in the questionnaire to assess the frequency of this practice among a larger pool of engineers. For example:
\\
\\
``Can you describe some cases where you would consider reference songs from another Genre?''
\\ 
\\
``Consider a situation where you like how an instrument is processed in a song from another genre [not the genre of the song you are working on] and you want to adapt this to the instrument of your song.''
\\
\\
By including these questions, we aimed to gather quantifiable data and gain a broader perspective on the communication and collaboration dynamics between clients and mixing engineers in the song-mixing process. The questionnaire took into account a diverse range of mixing engineers to ensure a more comprehensive understanding of the topic, incorporating insights and themes derived from the initial interviews.

\begin{figure}
    \centering
    \includegraphics[scale = 0.049]{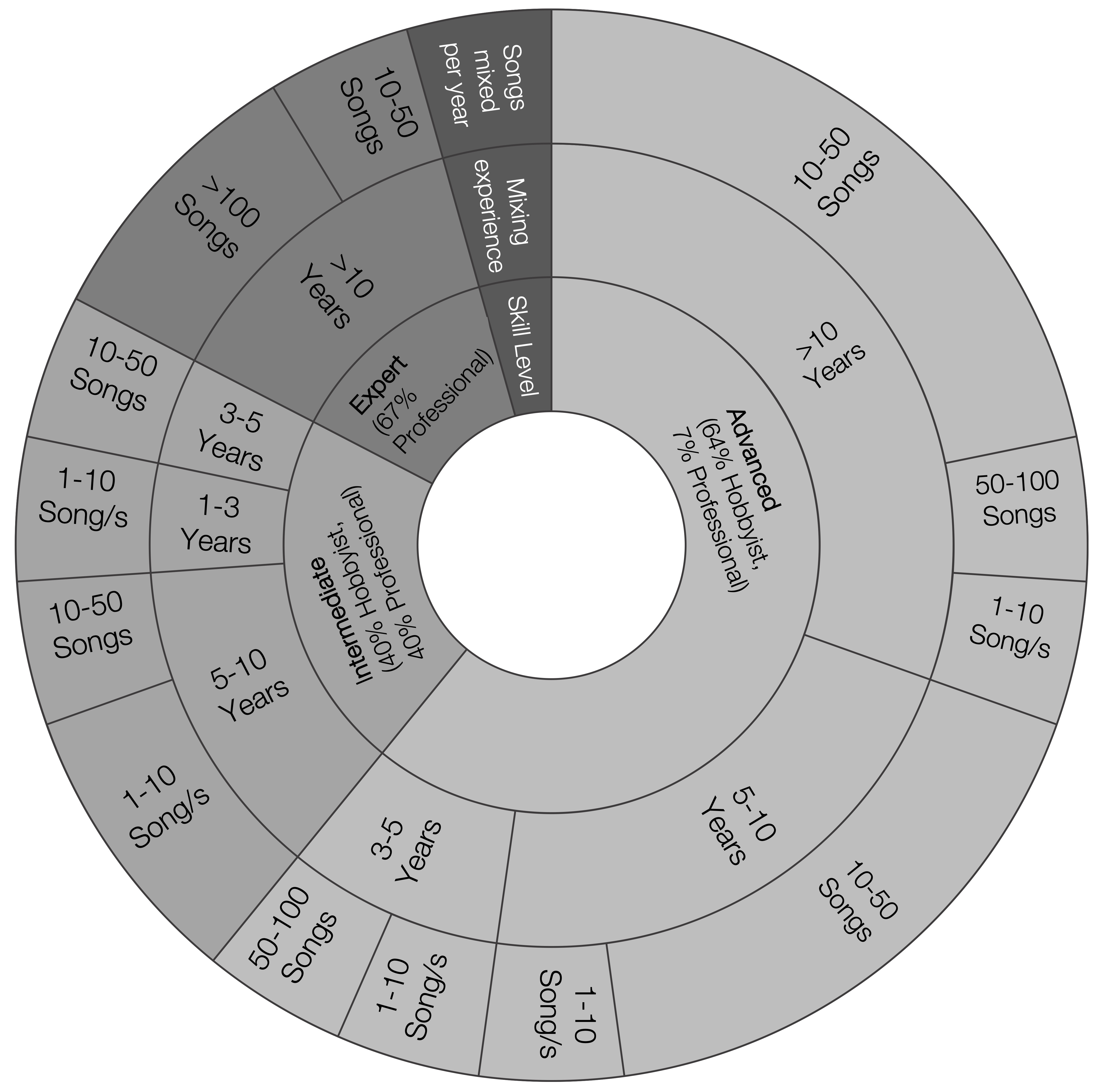}
    \caption{Mixing Engineer's profile for Phase 2 of the study}
    \label{fig:profile_p2}
\end{figure}

\subsubsection{Questionnaire Structure}
To ensure comparability across participants, we structured the questionnaire in a standardised manner~\cite{brinkmann2014unstructured}. It consisted of three sections. The first section collected general information about participants' musical practices, expertise and skills. The second section focused on standard practices, client interaction, and the use of demo mixes and reference songs in the mixing workflow. The third section explored the participants' understanding of the information provided by reference songs and demo mixes, as well as the role of creativity in mixing practice. The questionnaire included 33 questions, which were mostly multiple-choice or multi-select, with some optional long-answer questions. It was hosted online using Microsoft Forms and took approximately 15-20 minutes to complete.

\subsubsection{Participant Profile}
The questionnaire was distributed among professional mixing engineers within our personal network. A total of 22 participants with varied experiences took part in the study. \autoref{fig:profile_p2} illustrates the experience levels of the participants, with all of them having more than 3 years of production experience. The majority of respondents identified their mixing skills as advanced. They worked on a range of 1-10 to over 100 songs per year. The participants consisted of professional or pro-am mixing engineers who worked on their own music as well as projects for other artists. The participants were based out of Asia, Europe, and America. About 95\% of the participants were male. About 50\% of them worked in the pop, rock, electronic, and metal genres, however, we also had engineers who specialised in classical, hip-hop, jazz, folk, and world music in decreasing order. 
\begin{figure}[t]
    \centering
    \includegraphics[scale = 0.17]{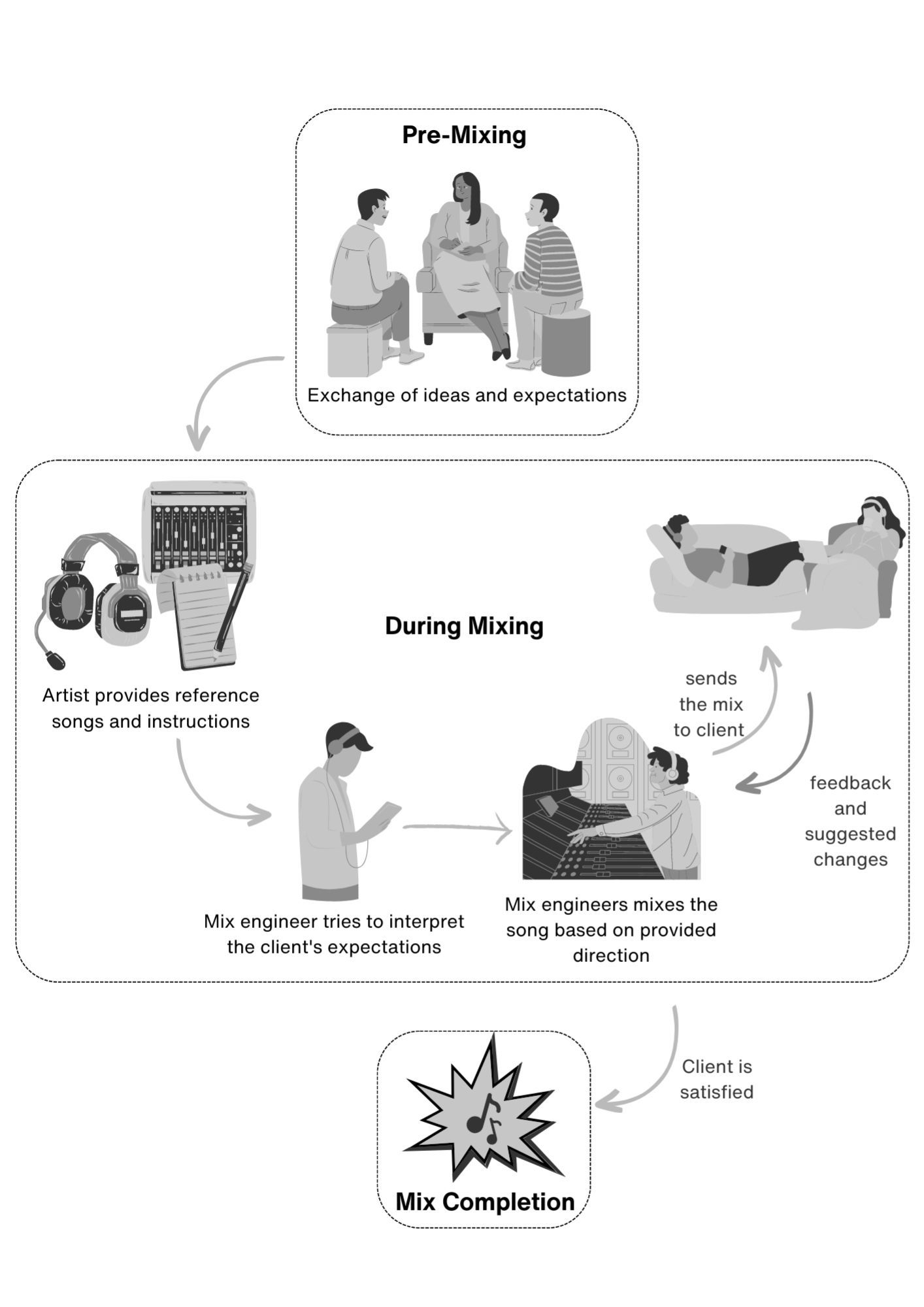}
    \caption{Flow of communication and collaboration during the mixing process}
    \label{fig:graphical_abstract}
\end{figure}
\subsubsection{Data Analysis}
For the questionnaire data, a mixed methods approach was employed due to the nature of the data collected. Qualitative data was processed using Thematic Analysis~\cite{braun2012thematic, braun2006using}, while quantitative data was analysed using descriptive statistical methods in Microsoft Excel. This approach provided a comprehensive analysis of both qualitative and quantitative aspects of the data.

\section{Results}
The following section presents results from both study phases, categorised into three stages: pre-mixing, during mixing, and mix completion. The findings from both the interviews and the questionnaire are discussed together, encompassing both quantitative data from the questionnaire and qualitative insights from the interviews. An overview of this is shown in \autoref{fig:graphical_abstract}. The results of our research indicate a strong correlation between effective communication and collaboration and the quality and alignment of the final product. 

\subsection{Pre-Mixing}
Effective communication with the client is an important part of the mixing process. It is important for the mix engineer to understand the client's expectations and preferences, and to work with them to achieve the desired sound. When asked what mixing means to them, engineers said, 
\\
\textit{``It's all a thing about communication and trying to understand what the client wants.''
\\
``Their hopes, I would say.''}
\\
Engineers often set up an initial consultation session with the artist where they try to get a sense of the artist's vision for the mix. The session involves discussions about the sound of individual elements of the mix as well as the mix as a whole.

\begin{figure}[th]
    \centering
    \includegraphics[scale = 0.059]{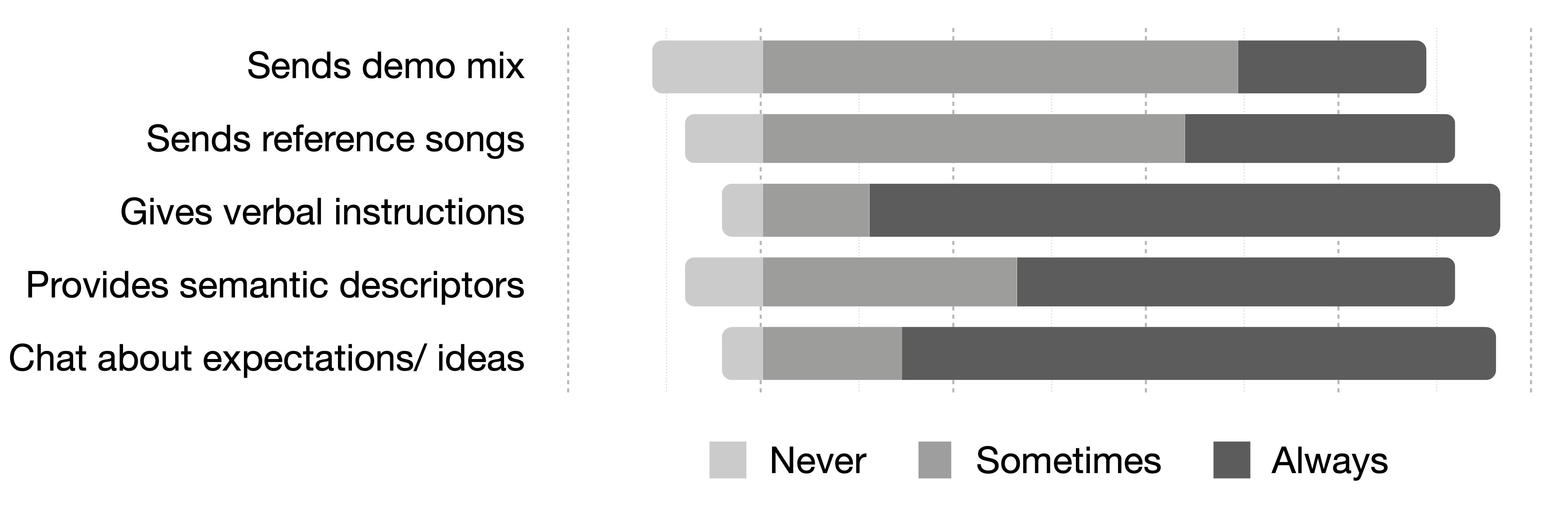}
    \caption{Ways in which expectations about the mix are conveyed}
    \label{fig:expectation}
    \end{figure}

As seen in \autoref{fig:expectation}, the artist uses different mediums to convey their expectations. The artist may describe the desired sound using verbal instructions, and semantic terms, provide reference songs, and offer a rough mix (also called a demo mix) as a guide. Semantics are words, linguistic units, or phrases applied to capture the meaning and interpretation of various sonic elements of the mix. Previous work has explored the role of semantics in music production and how it is used to describe the sound and expected transformation for the sound~\cite{stables2014safe, stables2016semantic, moffat2022semantic}.

\textit{``They give you some description of how they would like the mix to be. Yeah, sometimes it's just a verbal reference in terms of you know, hey have you heard that song, maybe something on that space something in this space, or it could just be like how we are having a chat … I need it very bright. So, usually, we tend to think, then I try to understand from him what he means by bright. Usually, it tends to be like he needs a lot of highs in the sound. You know high frequencies.''} 

The use of the term ``bright'' to describe the sound is a common example of how semantics can be applied to convey specific characteristics. In this context, when the client describes the sound as ``bright,'' they are likely referring to a sonic quality that emphasizes higher frequencies or has a sense of clarity and brilliance. 

Ultimately, the mix engineer should work closely with the artist to ensure that the final mix meets their expectations and achieves the desired sound. 

\subsection{During Mixing}
Mixing is a process of carving the sound of the final mix that emotes and tells the intended story. The mix engineer begins by loading the raw tracks or stems into the DAW and organising them in a way that is easy to navigate and understand. This may involve labelling the tracks and grouping similar ones together. Next, the engineer listens to the project and takes note of any issues that need to be addressed. They may listen to the project multiple times, as well as the rough mix and reference songs provided by the client to get a sense of the desired direction for the mix. After this initial preparation, the engineer uses the gathered direction to make decisions on the technical aspects of mixing, such as gain staging, panning, and fixing any frequency masking and noise or clicks. They then proceed to add more creative elements such as reverb, delay, and distortion. All of these decisions are made in alignment with the vision for the mix. 
% \usepackage{multirow}
% \begin{table}
% \centering
% \renewcommand{\arraystretch}{1.5}
% \begin{tabular}{p{0.3\linewidth}|p{0.2\linewidth}|p{0.15\linewidth}|p{0.18\linewidth}}
% \hline
% \textbf{Directors} & \textbf{Status} & \textbf{Usage (in\%)} & \textbf{Receiving from client (in\%)} \\ \hline
% \multirow{3}{*}{\textbf{Demo Mix}}        & Yes       & 27 & 24 \\ \cline{2-4} 
%                                          & Sometimes & 50 & 38 \\ \cline{2-4} 
%                                          & No        & 23 & 38 \\ \hline
% \multirow{3}{*}{\textbf{Reference Songs}} & Yes       & 45 & 14 \\ \cline{2-4} 
%                                          & Sometimes & 45 & 38 \\ \cline{2-4} 
%                                          & No        & 9  & 48 \\ \hline
% \end{tabular}
% \caption{Likelihood of usage of various directors in the mixing process.}
% \label{table:directors}
% \end{table}
\begin{figure}
   \centering
   \includegraphics[scale = 0.065]{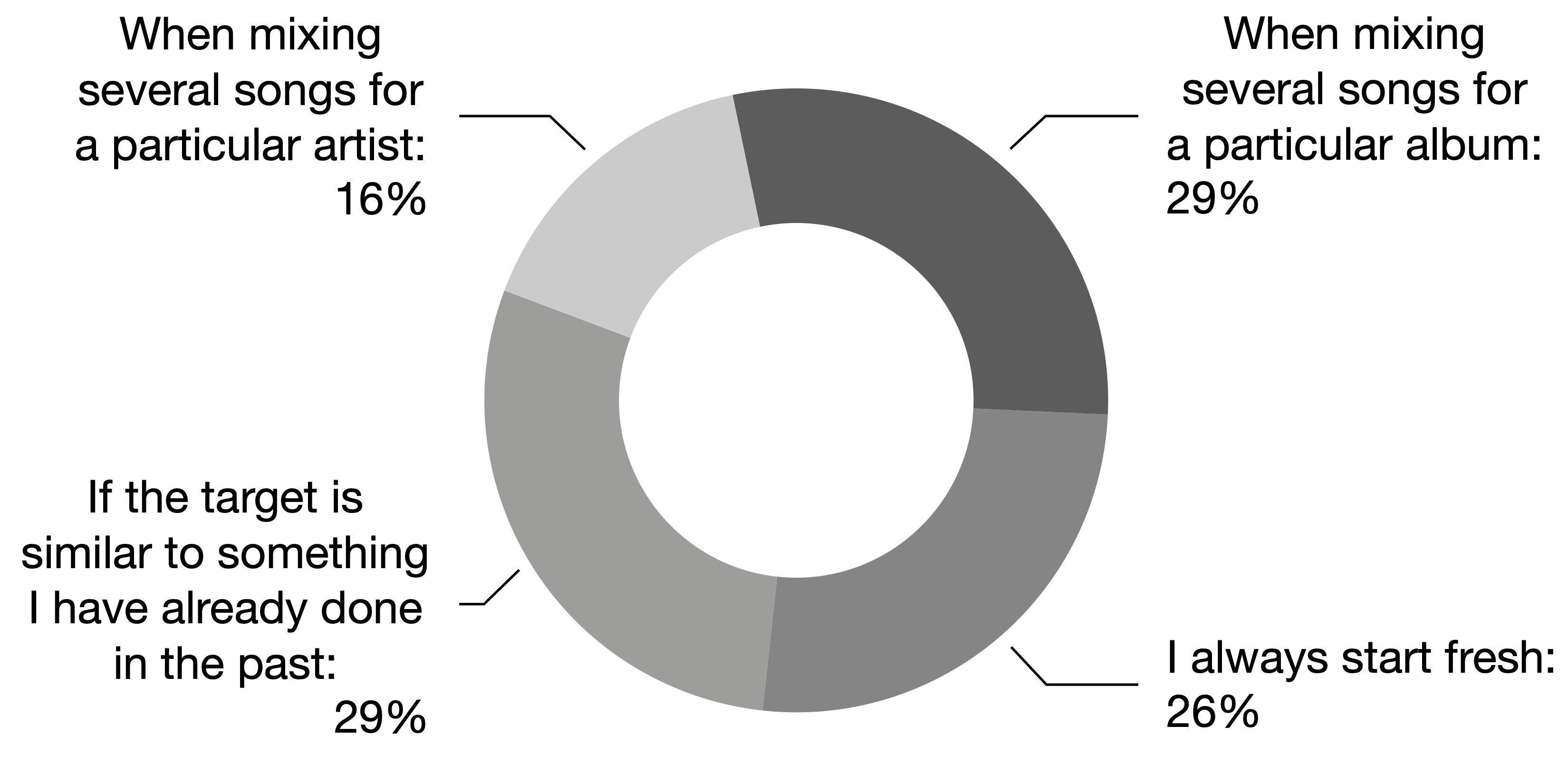}
   \caption{Situations when engineers import settings from the previous project}
   \label{fig:importsettings}
\end{figure}
\subsubsection{Inspiration from Previous Work}
Mix engineers often have their own techniques and approaches to mixing. \autoref{fig:importsettings} shows that some engineers prefer to start a project from scratch, while others find it useful to borrow ideas or settings from past projects.

This is true in three cases: 
\begin{enumerate}
    \item Artist-specific sonic preference: Some artists have a preference for a certain sonic identity for elements in their mix. This could be unique to the artist’s identity. In these cases, the artist might request the engineer to take inspiration from previous songs/projects of the artist. 
    \\
    \textit{``For example, I was mixing a song for somebody and that artist really liked the way her voice sounded in one particular mix. Next time a very different song was done but, she wanted the vocals to sound just the way it was before. So, I did turn back to what I had done and you know present it, and that worked for that artist.''}
    \\
    This highlights the importance of understanding the client's preferences and using past projects as a reference to achieve their desired sound. 
    \\
    \item Album-specific sonic preference: It is a practice to maintain the cohesiveness of sound across an album. Hence, if an engineer is working on all the songs in an album, they generally establish a sound for a song on the album and then take inspiration from that for mixing the rest of the songs. 
    \\
    \textit{``It also depends, if I'm mixing just one song or a whole album and if I'm mixing a whole album versus just like one song, I will typically ask the artist for a song, that they'd like me to start with to establish the overall sonic characteristics of the album. … their overall concept for the album needs to be cohesive. So, I will, ...mix one song, make sure it's going in the right direction, and then use that to lead me into every other song and start to borrow from that first mix and bring things over other mixes and change them accordingly.''}
    \\
    Importing settings from other tracks on the album can also be a time-saving technique, especially if the tracks were recorded in the same acoustic space. 
    \\
    \textit{``So quite often when you're mixing a whole album, the first thing you do is if you got the same drum kit and it was all recorded in the same room. Yeah, just start with the same drum sound on every song and then do the creative part.''}
    \\
    \item Processing-specific sonic preference: Often mix engineers borrow sound processing ideas from previous projects when they feel that the current mixing project can benefit from a starting point as the sonic sound goal is similar to something they have done previously. 
\end{enumerate}

It is important to remember, however, that every song is unique and may require its own approach and treatment.
In the coming subsections, we will elaborate further on how elements like demo mixes and reference songs are used to aid the process of mixing a song. 

\begin{table}[th]
\centering
\renewcommand{\arraystretch}{1.3}
\begin{tabular}{p{0.3\linewidth}|p{0.2\linewidth} p{0.15\linewidth} p{0.16\linewidth}}
\hline
\textbf{Directors} & \textbf{Status} & \textbf{Usage (in\%)} & \textbf{From client (in\%)} \\ \hline
\multirow{3}{*}{\textbf{Demo Mix}}        & Yes       & 27 & 24 \\  
                                          & Sometimes & 50 & 38 \\  
                                          & No        & 23 & 38 \\ \hline
\multirow{3}{*}{\textbf{Reference Songs}} & Yes       & 45 & 14 \\ 
                                          & Sometimes & 45 & 38 \\ 
                                          & No        & 9  & 48 \\ \hline
\end{tabular}
\caption{Likelihood of usage of various directors in the mixing process.}
\label{table:directors}
\end{table}
\subsubsection{Demo Mix (Rough Mix)}
\begin{figure}[th]
    \centering  
    \includegraphics[scale = 0.058]{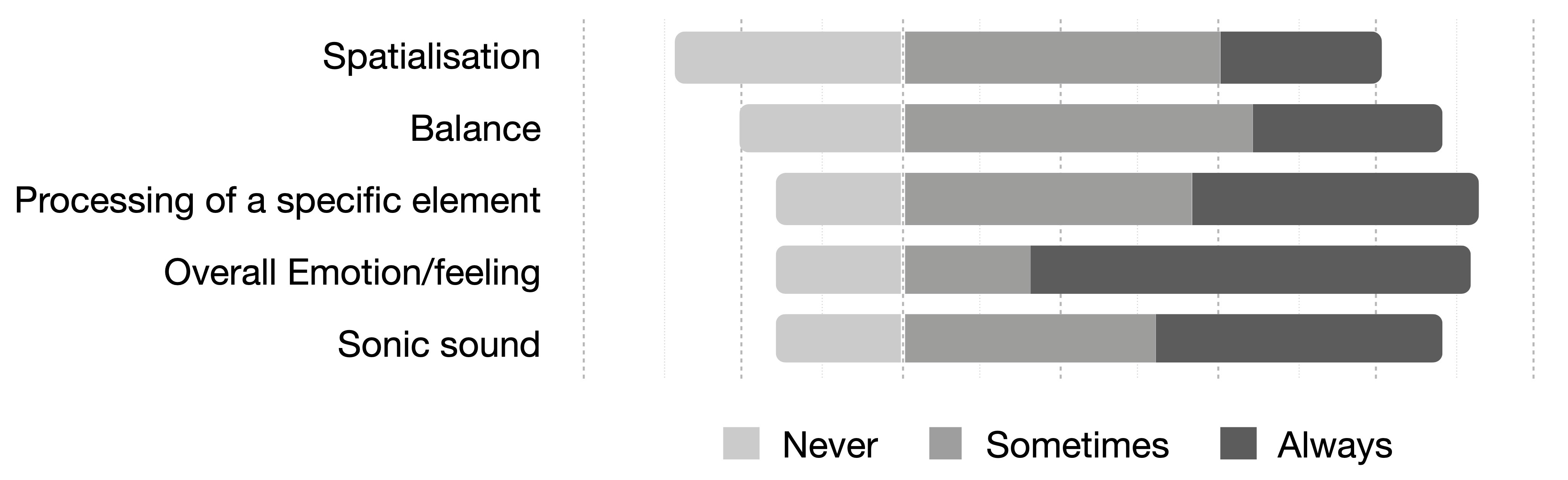}
    \caption{Information derived from Demo mixes}
    \label{fig:infromationfromdemo}
\end{figure}

Rough mix or demo mix are terms used to describe an unfinished coarse mix of a song that is completed by adjusting gain and pan levels and basic processing on some of the elements. Our studies (\autoref{table:directors}) showed that about 77\% of the participant mix engineers were likely to use demo mixes and more than 50\% of the time, they received a rough mix from the client. It is typically used to give the mix engineer an idea of the artist’s vision for the mix and can be helpful in providing a sense of direction for the mix as also presented in \autoref{fig:infromationfromdemo}

\textit{``For me a demo, it's not finished. It's more like this is the start of how we're going to.''  
\\
``But a demo song as well could be really helpful for me to understand where the thing should go.''}
\\
However, it's important to note that the demo mix is often not perfect and may require adjustments to get to the final mix.
\\
\textit{``In 99.9\% of the cases, it's really bad. But most of the clients would give you some sort of a demo or something that they make themselves so that you get a direction of what they want to do.''}
\\
The rough mix is sometimes done by the recording engineer, producer or the artist themselves. However, demo mixes are often rough and may not be fully representative of the desired sound, so it is important for the mix engineer to have a conversation with the client to understand their expectations and preferences. The client might ask the engineer to stick with the balance, panning, or processing on certain instruments or elements as shown in \autoref{fig:infromationfromdemo}. 
\\

\textit{``Usually I'm mixing from a rough mix as well. They do a rough mix and I usually ask them the question very early. How important is this rough mix to you? Is this part of your vision or do you want me to go in the opposite direction and do something.''}
\\
\textit{``I will speak with the composer or the client you know what are you wanting me to listen to from the demo mix? So, sometimes they will tell me that this is the producer's mix, so don't change the balance of it. I want the balance to be the same.''}
\\
The artist may ask the mix engineer to focus on certain elements of the rough mix, such as balance or panning, or to preserve certain elements of the rough mix.

\subsubsection{Reference Songs}
\begin{figure}[t]
    \centering
    \includegraphics[scale = 0.046]{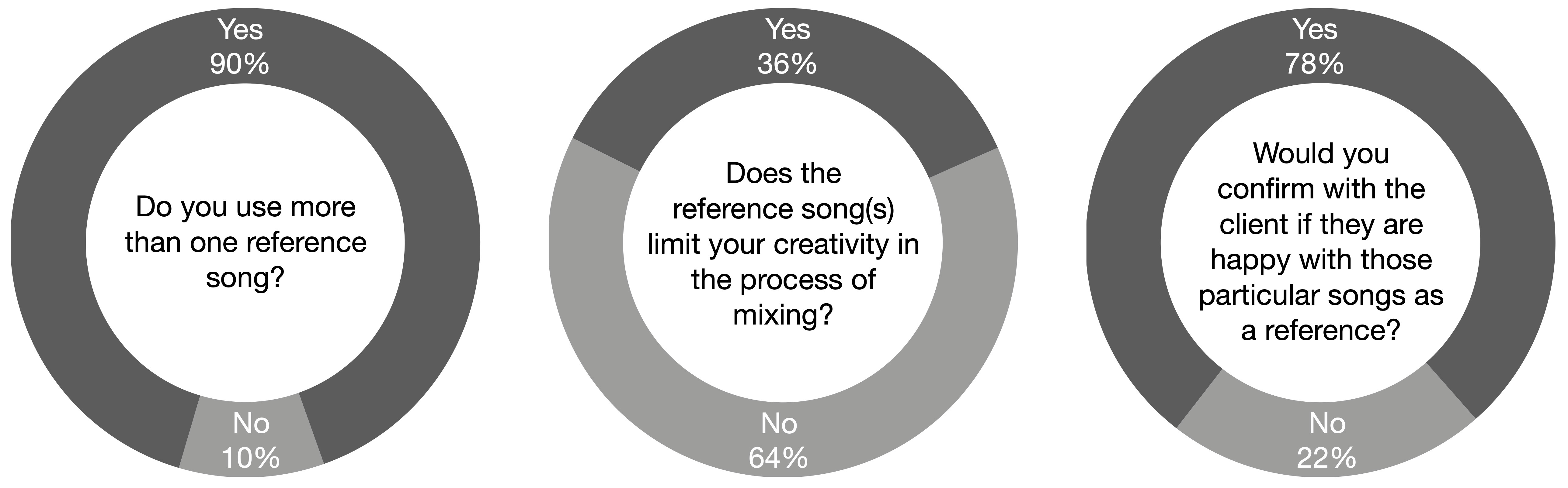}
    \caption{Influence of the reference song in the mixing process}
    \label{fig:yesno}
\end{figure}

Apart from demo mixes, most mix engineers often also rely on reference songs to guide them in the mixing process as shown in \autoref{table:directors}. A reference track is a song from another artist to use as a benchmark for varying elements in your own productions. Moreover, 90\% of the engineers confirmed that they use more than one reference song as seen from  \autoref{fig:yesno}.

\begin{figure}[th]
    \centering
    \includegraphics[scale = 0.059]{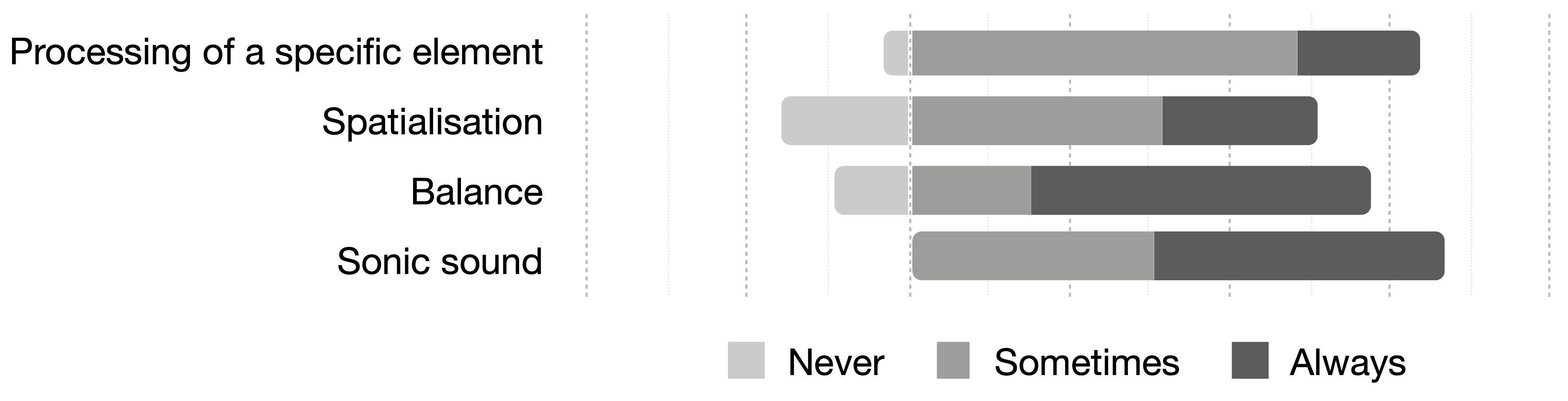}
    \caption{Information derived from the reference songs}
    \label{fig:refmixinfo}
\end{figure}

\begin{figure} 
    \centering
    \includegraphics[scale = 0.064]{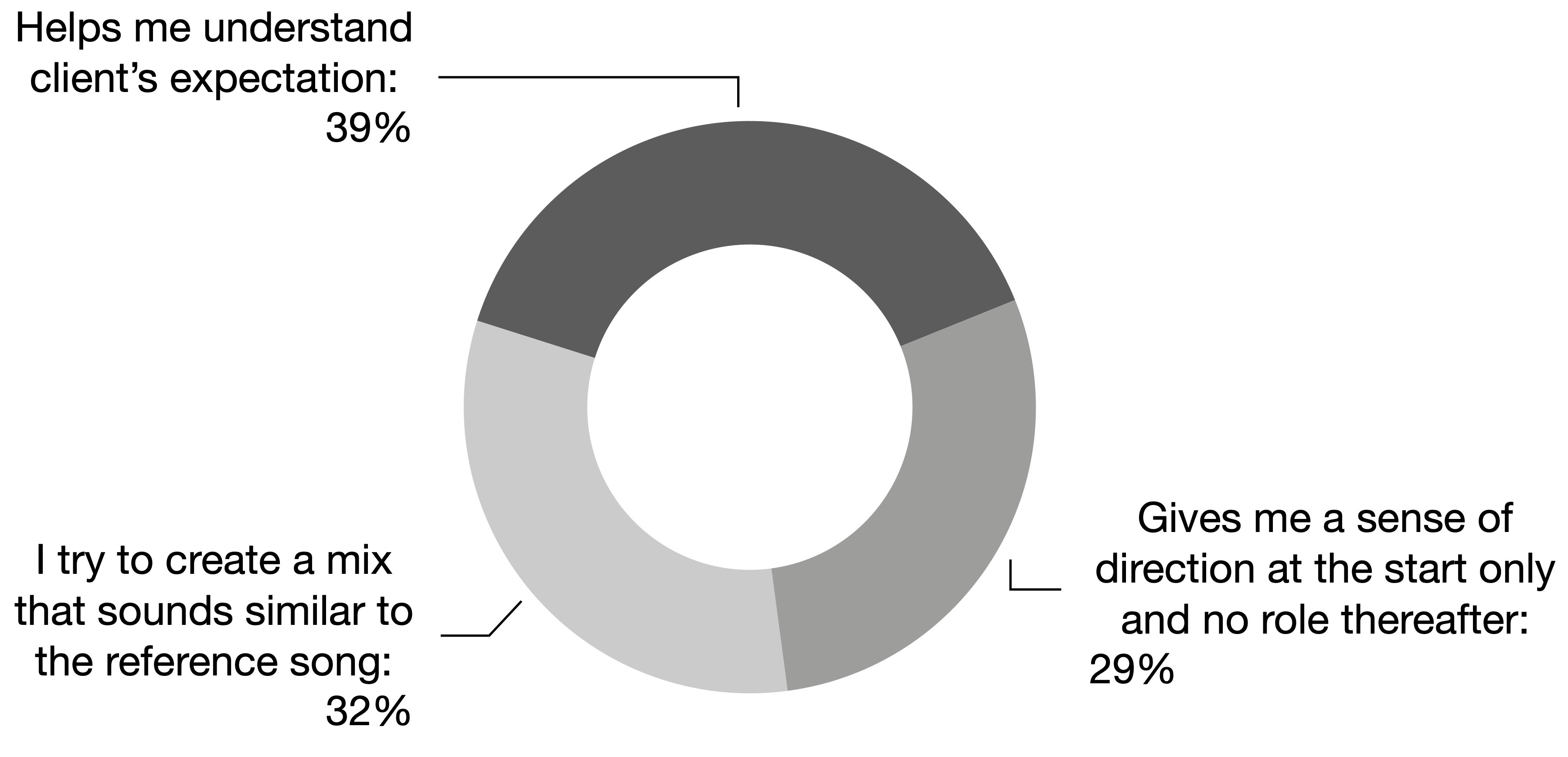}
    \caption{The use of reference songs in the mixing workflow}
    \label{fig:refinmix}
\end{figure}

\paragraph{Directions from reference song}
When asked what the use of reference songs is in the mixing workflow, engineers described (as seen in  \autoref{fig:refinmix}):
\begin{enumerate}
    \item Helped understand the client’s vision
    \item Get a sense of direction
    \item A pointer for the sound of the final mix: This may include understanding the balance, panorama, and processing of certain instruments as shown in \autoref{fig:refmixinfo}.
\end{enumerate}

\begin{figure}[t]
    \centering
    \includegraphics[scale = 0.065]{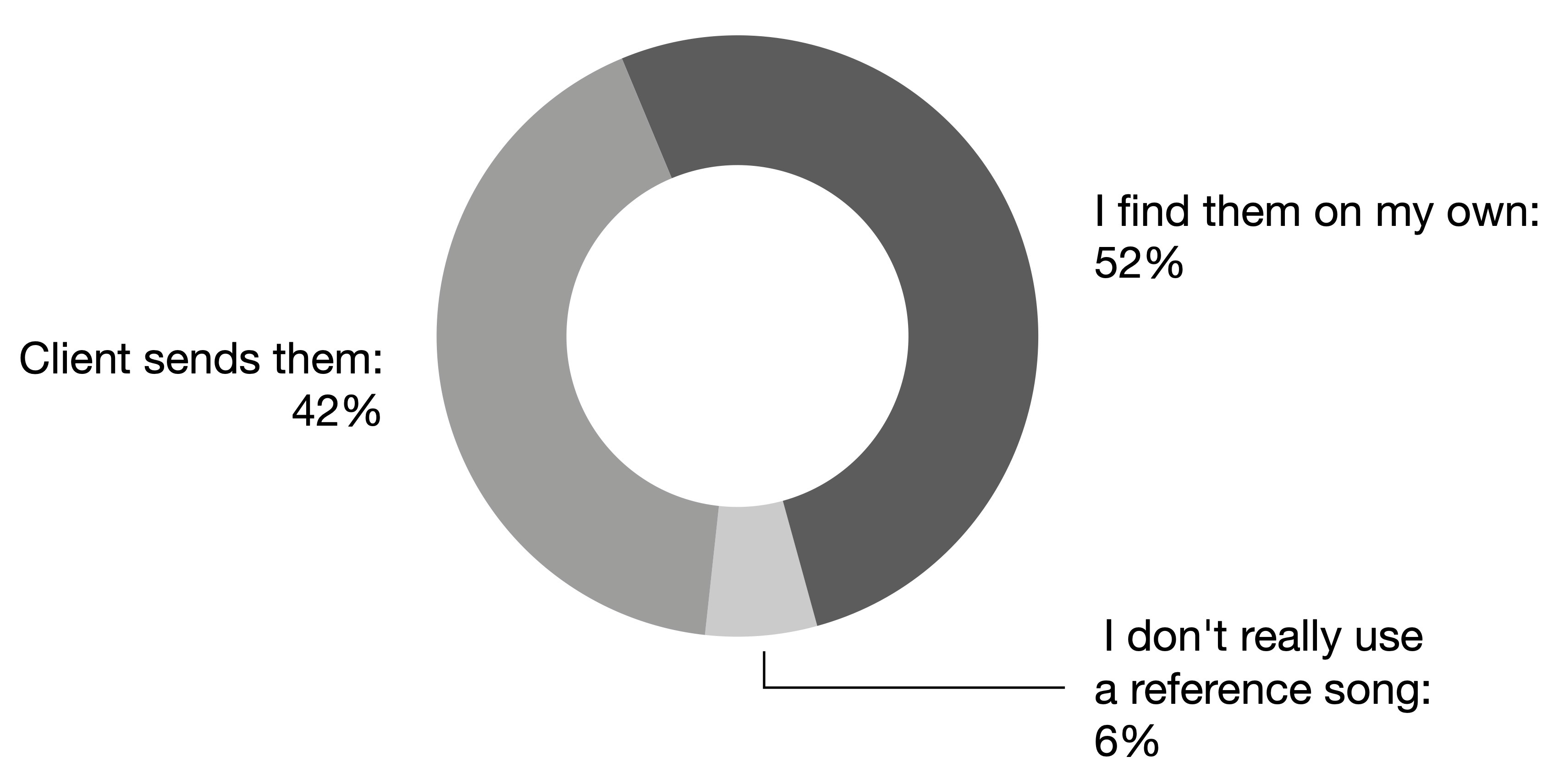}
    \caption{How do you find a reference song?}
    \label{fig:findref}
\end{figure}

Reference songs are commonly used by mix engineers as a direction for their work. 
\\
\textit{``Most often a reference song acts as a direction sign on the road. It helps me get a sense of expectation and direction.''}
\\
The reference song either brings out the same target emotion, has a similar target semantic descriptor (for example the song sounds warm, bright, etc), has the same target sonic and dynamic sound or has a similar compositional structure. In addition to providing a sense of direction, reference songs can also be used for A/B testing to see how far the engineer has progressed with the mix, and to help composers and artists communicate their vision to the engineer by presenting examples in the form of a reference song. However, it is important to note that while reference songs are often chosen for their ability to evoke a certain feeling or bring out a specific emotion, they are typically not used to replicate the mix exactly as seen from a strong disagreement in \autoref{fig:compare}. 
\textit{``I'm not specifically interested in my mix sounding exactly the same. I'm interested in whether my mix feels the same.''}
\paragraph{Finding a reference song}
We also induced that about 50\% of the time, clients send mixing engineers reference songs. As seen in \autoref{fig:findref}, about 42\% of the time, clients send reference songs, however in case the client does not, engineers find them on their own. Also, 78\% of the engineers confirmed that once they find reference songs, they pass it on to the client to confirm if the songs align with their vision as seen from \autoref{fig:yesno}. However, for certain purposes of usage like the ones described below, engineers may not get feedback from the client.
\\
\textit{``Depends on the purpose of the reference track. If the reference track is aimed at the overall sound and artistic vision of the artist i.e. a genre-specific reference track, then yes. If it's related to being able to use a reference track that I know has a well-balanced mix, then there's no need to discuss this with the artist as their judgment on that part of the process is not always necessary.''}
\\
\textit{``It depends on why I'm using the reference. For overall vibe, balance, etc. it's always good to get a client's input. More often than not I'm using a reference for some smaller detail (e.g. exactly how two sounds interact in the frequency spectrum), and in these cases, I don't bother getting a client's input.''}
%\begin{figure}[th]
%    \centering
%    \includegraphics[scale = 0.7]{figures/Q18.pdf}
%    \caption{If you find reference song/songs on your own, would you confirm with the client if they are happy with those particular songs as a reference? }
%    \label{fig:findref_2}
%\end{figure}
\paragraph{Choosing a reference song}
Choosing a reference song is an important part of the mixing process. Reference songs are generally chosen based on the purpose they intend to serve. 

\begin{enumerate}
     \item Director for the entire mix: Reference songs can serve as a guide for the engineer in order to achieve the desired sound and emotion for their mix. 
     \\
     \textit{``I'm definitely listening to the ambience and spatiality. I'm listening to compression. How punchy is it? Or is it quite a sort of mellow in the overall league? I'm not listening to specific EQs, but I'm listening to an overall EQ curve.''}
     \\
     The referenced song should either have a similar sound to the final mix, which could mean it has the same frequency profile, dynamics, and general processing, or brings out the same emotion.
     \\
     \textit{``It was you know very emotive and I had a reference in my mind that in this particular song, it should sound that breezy or that lush. So, I went back to that reference song in my head and looked it up and saw how is it sounding in that song and then I got a few clues and then I try to incorporate that in my song.''}
     \\
     \item Director for a particular element in the mix: Reference songs can give information about the processing or sound of one particular element or instrument in the mix. 
     \\
     \textit{``If someone sent you  a track it can mean so much because yeah, if you, if you're listening to something, then it's very personal on the things you're going to check out. I mean, some people are only going to do it for the vocals or for the whole song, or for structure. So, then I'm going to have to start a talk about it I think.''}
\end{enumerate}

A reference song is often from the same genre and sounds similar to the final expected mix, however, it is possible for it to be from another genre. This holds true in two situations: {\em i)} when the reference gives information about the processing of some specific element and not the mix as a whole, or {\em ii)} when the artist simply wants to evoke a certain emotion and is not concerned with genre. 
\\
\textit{``I always try to go to a song from the original genre or an album that reminds me of that song. I got this kind of blues song which was really warm but they didn't give me any reference. I was listening to the rough mix and I was like this reminds me of something. I went through my dance collection of CDs and I picked up a couple of songs and sent them to the artists saying this rough mix reminds me of these songs. Do you approve to use them as references?''}

\paragraph{Creativity}
The process of mixing music is technical as well as creative. In our studies, we asked engineers whether reference songs restrict their creativity, with the majority (about 64\% of the engineers) disagreeing, as indicated in \autoref{fig:yesno}. They emphasised that reference songs only provide a direction (a suggestion), however, the goal can be achieved in multiple ways. Reference songs provide a valuable means to grasp the client’s expectations. One engineer expressed that the sounds and instruments in their mix might be different from those in the reference, however, a reference might help them listen to the sound in the mix that they might have otherwise ignored. 
\\

\textit{``I would put like about 20-30\% of my own opinion into that. So, it doesn't really deviate too much from the actual reference, but then at the same time there's a bit of a character that is not the same as the thing that they bring in the reference song.''}
\\
\textit{``For helping me inform specific decisions and resolve uncertainties and not a creative limitation.''}

\paragraph{Feedback}
The process of mixing requires a continuous dialogue between the mix engineer and the client to ensure that the final product meets the client's expectations. One of the crucial elements of this dialogue is the feedback that the client provides on the mix in progress. As the mix engineer works on the track, they send different versions of the mix to the client for their review, making necessary adjustments and tweaks based on the client's feedback. This iterative process allows the engineer to make sure that the mix is on the right track, and that the client is happy with the direction it is heading.
\\
\textit{``I feel like it's done and I can send it to the artist where if that's the first mix, I'm obviously happy with it if there is a revision.''}
\begin{figure} 
    \centering
    \includegraphics[scale = 0.065]{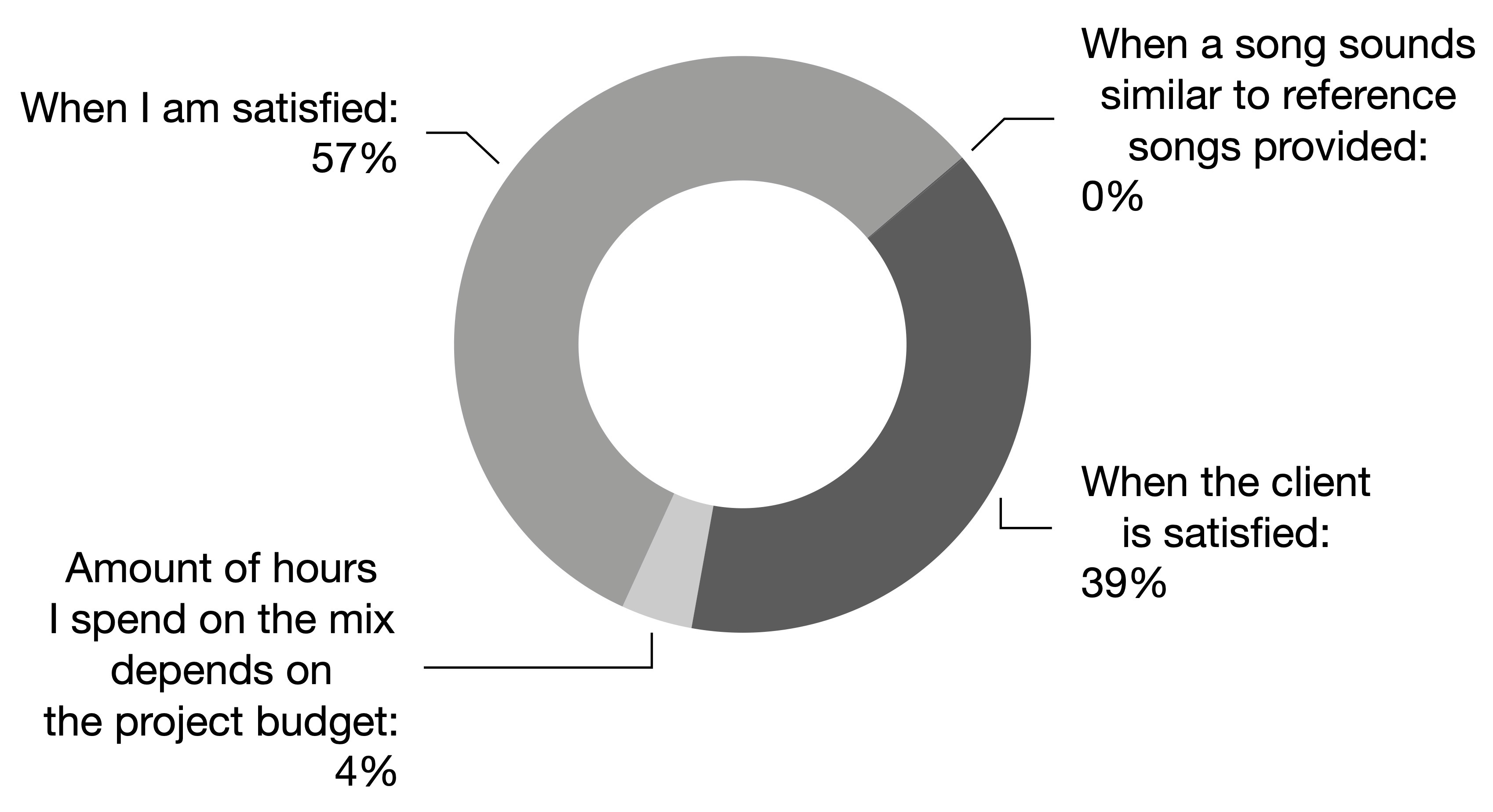}
    \caption{When do you know your mix is ready?}
    \label{fig:mixready}
\end{figure}
\subsection{Mix Completion}
Engineers agreed that the mix is ready whenever the mix sounds balanced, the engineer is happy, and the client is satisfied as shown in \autoref{fig:mixready}. 

\begin{figure}[t]
    \centering
    \includegraphics[scale = 0.065]{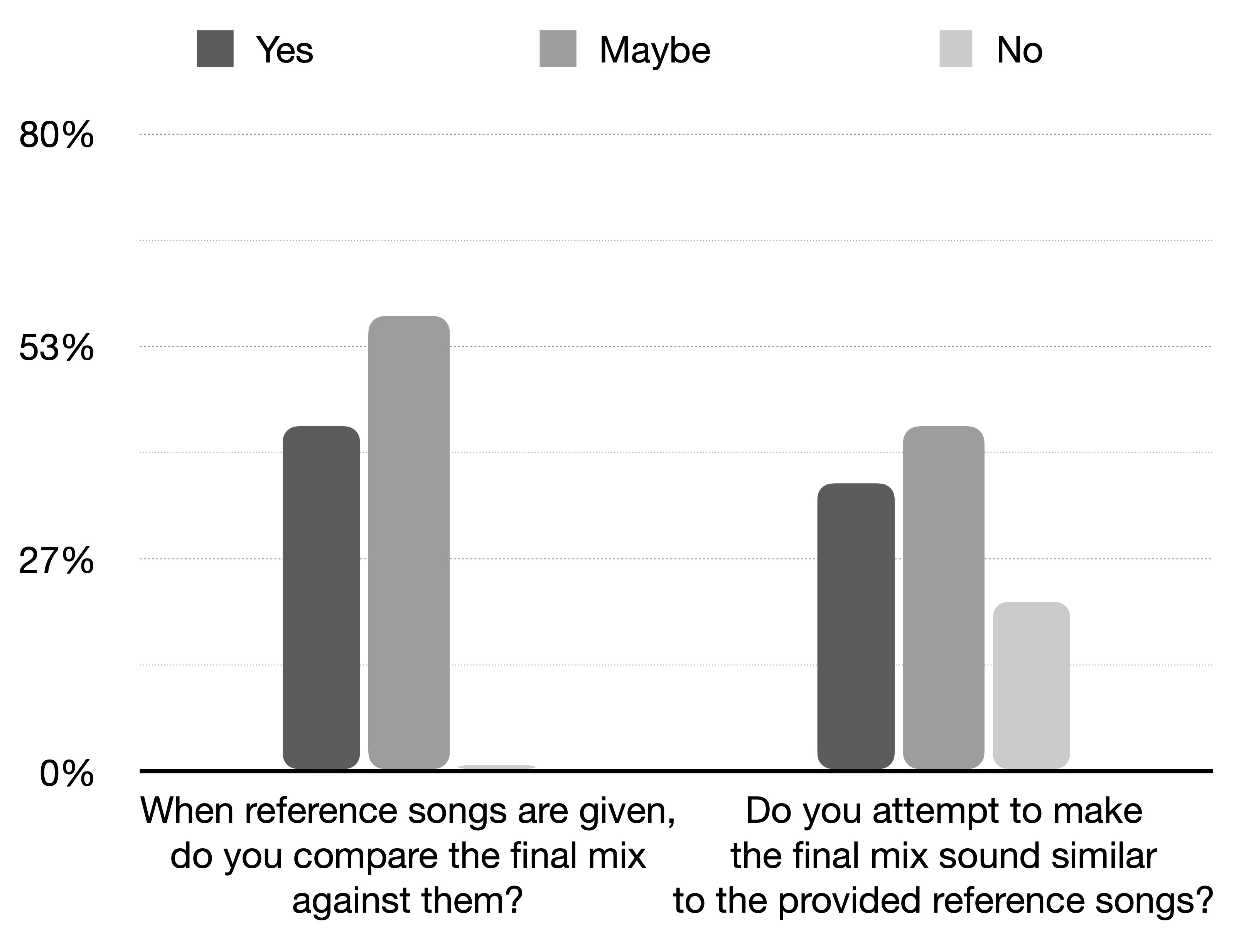}
    \caption{Comparing reference songs against mix}
    \label{fig:compare}
\end{figure}

\textit{``Honestly, I just hit the play button and just sit back and I am looking for anything that would surprise me and that's a negative thing. I mean, if I just sit back and listen to it and … it just sounds unnatural or something like perhaps my attention …I know that it's not finished.''}
\\
\textit{``I think it is done when the client says it's done because in your head it's never done. That's my experience. But, yes uh there comes a time a point in the mix where you're really happy.''}
\\
Engineers confirmed that often they compare the final mix against the given reference songs. Our studies also show that most often engineers try to create a mix that either sounds similar to the reference songs or evokes similar feelings as shown in \autoref{fig:compare}. 

\textit{``I feel it's like a middle point between the two like obviously when I feel like I'm happy with what I've done and I've got a reference I tried to compare it to the reference and if I feel what I have done is totally different, I'm like okay and I try to make it a bit closer''}
\\
\textit{``I'm not specifically interested in my mix sounding exactly the same. I'm interested in whether my mix feels the same.''}
\\
\textit{``I mean when it sounds good to me, generally balanced, and it has enough similarity to that reference.''}
\\
It is important to note that there isn't a one-size-fits-all answer to when a mix is ready. The decision ultimately depends on the individual engineer and the specific project they are working on. Some may rely heavily on reference songs to guide their mixing process, while others may be more focused on creating a mix that evokes the desired emotion or feeling. The mix is generally considered ready when it sounds smooth and balanced and evokes the same feeling as the reference song. Ultimately, a mix is ready when it meets the expectations of the engineer and the client and sounds good on a variety of playback devices.
%still working on this part. need to add references as well. 
\section{Discussion and Conclusion}
Our investigation highlights the crucial role of communication and collaboration in the mixing workflow. As mixing is a context-dependent task, it is essential for the mixing engineer to understand the artist's vision for the song and make decisions aligned with it to create the desired mix. The artist provides various mediums of information, including verbal instructions, semantic terms, reference songs and demo mixes, to convey their expectations. These directives provide a sense of direction to the mixing engineer, facilitating their decision-making process.

Our findings also shed light on the information that mixing engineers gather from each of these different directives, underscoring the dynamic nature of collaboration in the mixing workflow. Previous research has emphasised the importance of collaborative and co-creative smart mixing systems that promote interaction, collaboration, and trust among professionals ~\cite{vanka2023adoption, tsiros2020towards, lefford2021context}). This study further underscores the significance of incorporating context into the mixing workflow and presents a preliminary approach for designing such systems.

Context plays a crucial role in shaping mixing decisions and perceptions, encompassing factors such as style, genre, knowledge, time and technology. Moreover, context evolves dynamically throughout the song and the mixing process ~\cite{lefford2021context}). Our research demonstrates that context in music mixing can be incorporated through various means, including reference songs, demo mixes, verbal instructions and semantic terms. However, each of these directives presents its own set of challenges.

Previous studies on the utilisation of semantics to define mixing expectations have highlighted a significant challenge in generalisation, as there is no universally acknowledged and agreed-upon interpretation for specific semantic terms ~\cite{moffat2022semantic}. Additionally, the semantic terms may hold varied meanings across genres, production settings, and geographical regions~\cite{stables2014safe}. Verbal instructions provided in written form also require further examination. The complex nature of the mixing process may not be fully captured by a few words, making it difficult to represent the vision of the mix accurately.

In our work, we conducted a detailed investigation into the use of reference songs and demo mixes as mediums of context. The results demonstrate that reference songs have the ability to effectively convey the abstract nature of mixing style, capturing the desired emotions and sonic qualities sought by the artist. Similar findings have been observed in the music recording process regarding the use of reference songs. Ongoing research exploring the use of reference sound and reference songs for audio effects ~\cite{steinmetz2022style} and mixing style transfer ~\cite{mst_2021, koo2022music} has shown promising success in capturing context and the user's intentions.

The future success of AI in music production depends on the ability of these systems to support interaction. This interaction can be divided into two phases: before the mixing process and after the mixing process. An AI-based mixing system that enables smart mixing needs to be provided with context on the input side. This context can be provided in the form of reference songs, demo mixes, verbal instructions or semantic descriptors. Once the context is provided, the system should be able to interpolate and generate a mix that represents an understanding of the context. Furthermore, on the output side, the system should allow for fine-tuning the mix to achieve the desired sound. This is akin to iterating based on feedback provided in the collaborative dynamics of mixing. The system should either provide means for alteration and fine-tuning of the output or offer a way to receive feedback to further improve the result in the desired direction. 

In future work, our focus will be on investigating both subjective and objective evaluation methods to assess the system's effectiveness in capturing and incorporating contextual information provided into the output. Additionally, we aim to delve deeper into certain concepts that emerged in this study but were not extensively explored. One such concept is the notion of mixing style and its variations across different genres and artistic styles. 

Overall, our study underscores the importance of effective communication and collaboration in the mixing process. It provides insights into the different forms of communication employed by artists and the valuable information gained by mixing engineers from these channels. By understanding and incorporating these dynamics, future AI-based mixing systems can enhance the interaction and collaboration between artists and engineers, expectantly leading to improved outcomes in the music production process.

\section{ACKNOWLEDGMENT}
We express our sincere gratitude to the JAES reviewers for providing valuable feedback on our work.
We are truly appreciative of the contributions made by all the mixing engineers and participants who participated in the interviews and studies and shared their valuable knowledge with us. We extend our thanks to Steinberg's research and development team for their unwavering support and honest feedback throughout this project.

This work is funded and supported by UK Research and Innovation [grant number EP/S022694/1] and Steinberg Media Technologies GmbH under the AI and Music postgraduate research program at the Centre for Digital Music, QMUL.

% - If you are using BibTeX for references, you need these lines: 
%bibtex style changed. Might need to be changes back to JAES before submitting to JAES journal
%\bibliographystyle{jaes.bst}
\bibliographystyle{jaes.bst}
\bibliography{jaes.bib}

%\break

%Appendix
% \appendix

% One or more Appendices can appear after the references. An Appendix can present for example a mathematical derivation, pseudo-code, or other additional data, which is unsuitable to show in the body of the article.

%Biography
  \biography{Soumya Sai Vanka}{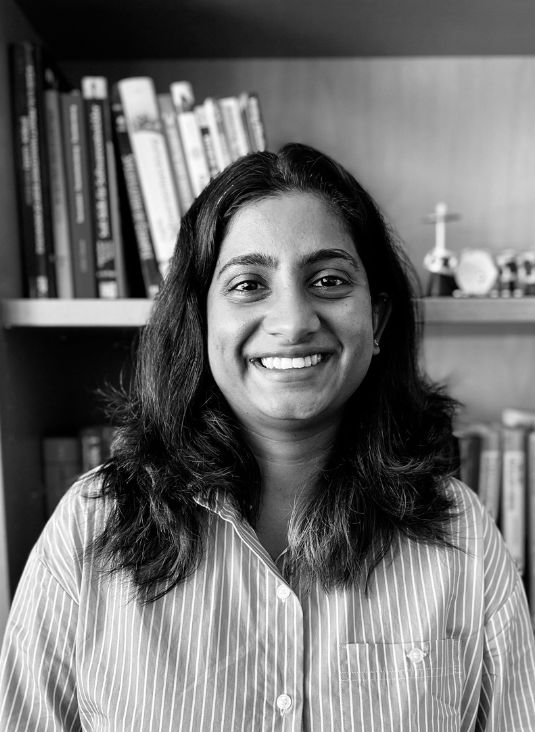}{Soumya Sai Vanka is a PhD student at Queen Mary University of London's Centre for Digital Music in the AI and Music program, collaborating with Steinberg Media Technologies GmbH. She holds a Gold medal in Physics (Bachelors[Hons]) from Sri Sathya Institute of Higher Learning, India, and a master's degree in Physics from Pondicherry University. Soumya is also an accomplished saxophonist who trained in music production and performed with various big bands. Her research focuses on expert-informed style transfer in music mixing, with interests in self-supervised machine learning, DDSP, intelligent music production, and human-centred design approaches.}

 \biography{Maryam Safi}{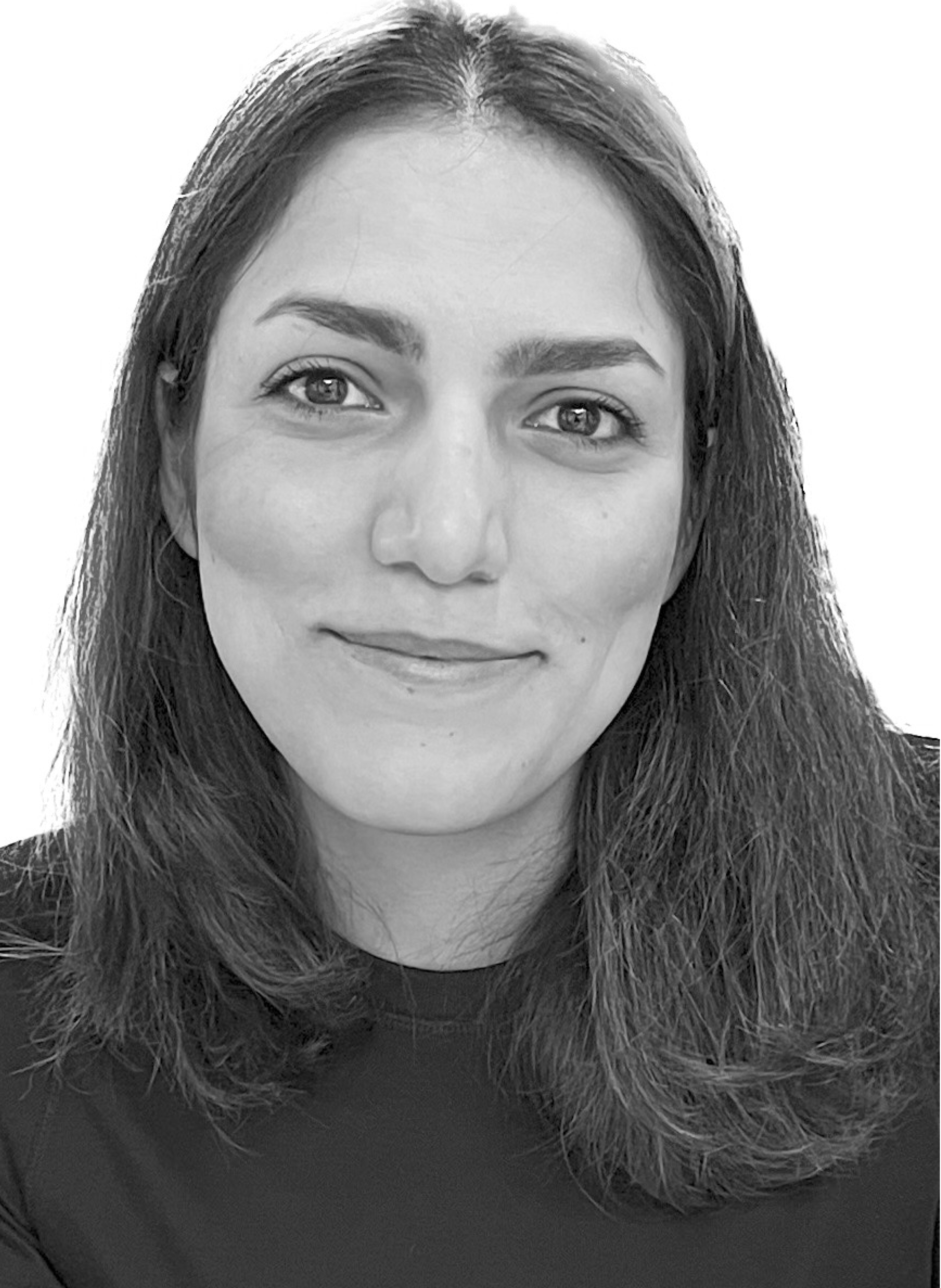}{Maryam Safi is an Audio Test Engineer in the R\&D team at Steinberg. Her work involves evaluating and supporting the attainment of superior quality standards in various aspects of DAW software development, such as existing audio effects plug-ins and prototypes of future products. With an academic foundation in statistics and media technologies, she is expanding her professional focus to audio quality in AI-based systems.
 She supports research in a wide range of topics, including Music AI and the integration of AI into audio plug-ins and mixing tools.
 Beyond her work, Maryam serves as the Vice-Chair of the AES Germany Section. In 2020 she was honoured with the Board of Governors award in recognition of exceptional leadership of the AES Student Delegate Assembly.}
 \biography{Jean-Baptiste Rolland}{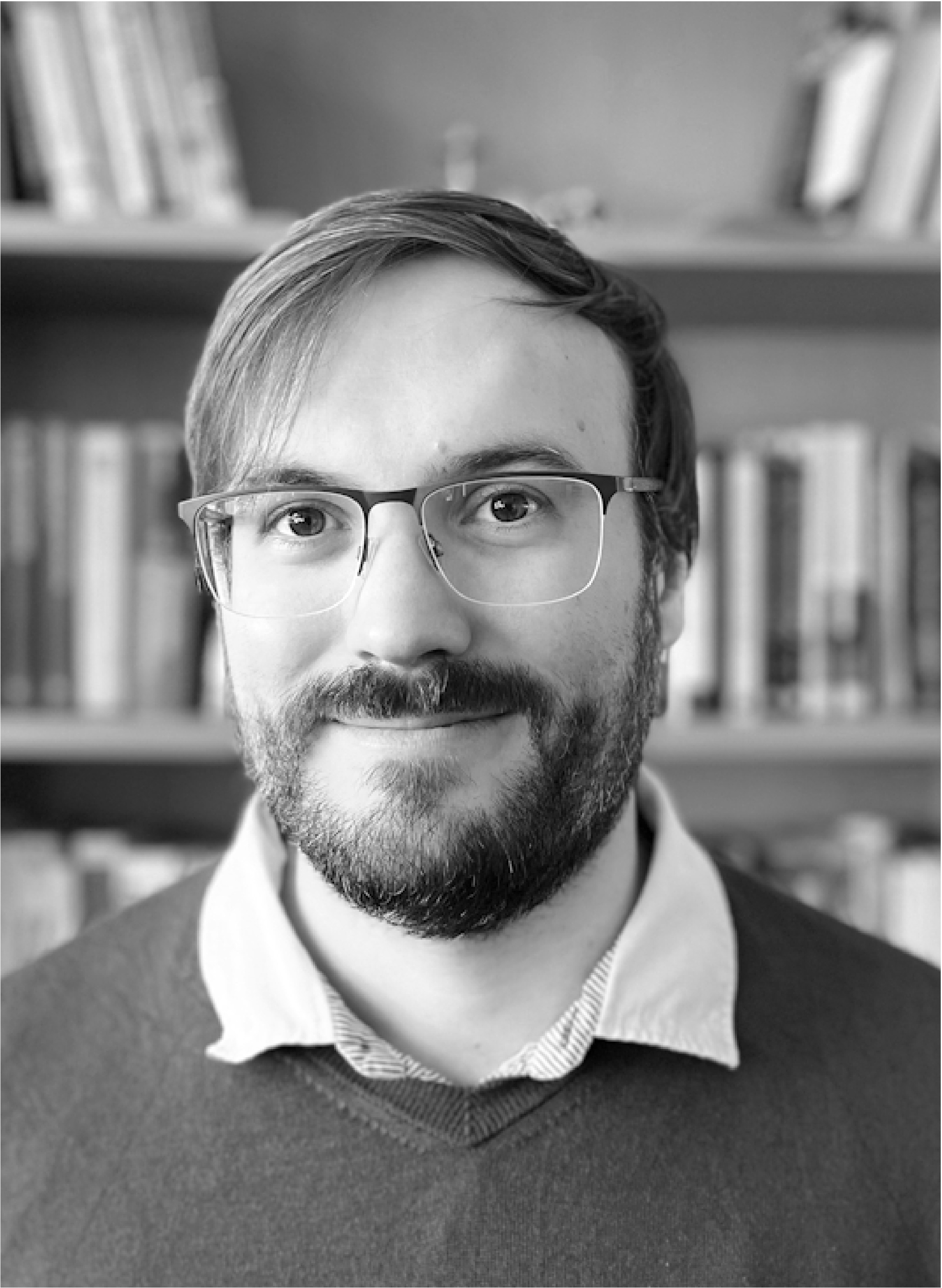}{Jean-Baptiste Rolland is a Senior Software Developer, contributing his expertise to Steinberg's Research Team. In this role, he focuses on practical research in audio analysis, bridging the gap between theory and application within Steinberg's product line. His work encompasses a broad array of fields, including Digital Signal Processing, Artificial Intelligence, and Software Engineering. He specializes in Music AI, Audio Analysis, Audio Restoration, and Music Information Retrieval, with his contributions resulting in features used, for example, in Cubase, Nuendo, and Wavelab. Jean-Baptiste is also actively engaged in academic collaborations and serves as an industry supervisor to students pursuing their academic work (PhD, master’s degrees), helping to shape the future talent in this field.}
 \biography{Gy\"orgy Fazekas}{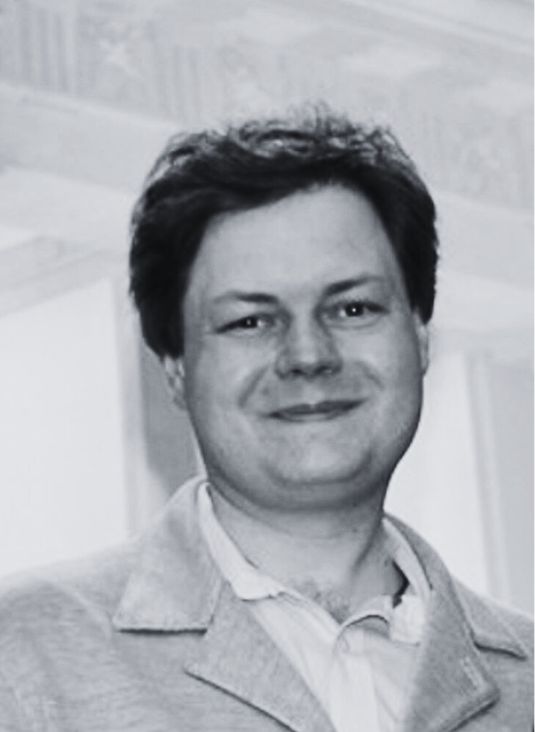}{Gy\"orgy Fazekas is a Senior Lecturer at the Center for Digital Music, Queen Mary University of London. He holds a BSc, MSc, and PhD degrees in Electronic Engineering. He is an investigator of UKRI’s £6.5M Centre for Doctoral Training in Artificial Intelligence and Music (AIM CDT) and he was QMUL’s Principal Investigator on the H2020funded Audio Commons project. He was general chair of ACM’s Audio Mostly 2017 and papers co-chair of the AES53rd International Conference on Semantic Audio and he received the Citation Award of the AES. He published over 180 papers in the fields of Music Information Retrieval, Semantic Web, Deep Learning, and Semantic Audio.}

\end{document}